\documentclass[showpacs,floatfix,superscriptaddress,showpacs,twocolumn,amssymb,amsfonts,prl,aps]{revtex4}
\usepackage[light,all,outline,bottomafter]{draftcopy}
\draftcopyName{DRAFT July 13, 2006}{140} \draftcopySetScale{70}

\usepackage{longtable,graphicx,epsfig,dcolumn}

\begin{document}
\bibliographystyle{revtex}
\title{Capillary Filling of Anodized Alumina Nanopore Arrays}

\author{Kyle~J.~Alvine}
\email{kalvine@post.harvard.edu} \altaffiliation[Current address:
]{National Institute of Standards and Technology (NIST),
Gaithersburg, MD (USA)}
 \affiliation{Division of Engineering and Applied Sciences,
Harvard University, Cambridge Massachusetts 02138 (USA)}

\author{Oleg~G.~Shpyrko}
\affiliation{Division of Engineering and Applied Sciences, Harvard
University, Cambridge Massachusetts 02138 (USA)}\affiliation{The
Center for Nanoscale Materials, Argonne National Laboratory,
Argonne, IL, 60439 (USA)}

\author{Peter~S.~Pershan}
\affiliation{Division of Engineering and Applied Sciences, Harvard
University, Cambridge Massachusetts 02138
(USA)}\affiliation{Department of Physics, Harvard University,
Cambridge Massachusetts 02138 (USA)}

\author{Kyusoon~Shin}
\altaffiliation[Current address:  ]{School of Chemical and
Biological Engineering, Seoul National University, Seoul (South
Korea)}
\author{Thomas~P.~Russell}
\affiliation{Department of Polymer Science and
Engineering,University of Massachusetts, Amherst Massachusetts
01003 (USA)}

\date{July 13, 2006}

\begin{abstract}

\def\baselinestretch{1}
\noindent  The filling behavior of a room temperature solvent,
perfluoro-methyl-cyclohexane, in $\sim$20~nm nanoporous alumina
membranes was investigated \emph{in-situ} with small angle x-ray
scattering. Adsorption in the pores was controlled reversibly by
varying the chemical potential between the sample and a liquid
reservoir via a thermal offset, $\Delta T$.  The system exhibited a
pronounced hysteretic capillary filling transition as liquid was
condensed into the nanopores.  These results are compared with the
Kelvin-Cohan prediction, with a prediction in which the effect of
the van der Waals potential is combined with the Derjaguin
approximation and also with recent predictions by Cole and Saam.
\end{abstract}


\maketitle

Nanoporous materials hold great potential in a wide variety of
nanotechnology applications such as: DNA
translocation~\cite{Meller01,Rabin05}, nanofluidic
transistors~\cite{Karnik05}, templates for nanoparticle
self-assembly~\cite{Alvine06,Lahav03}, catalysis~\cite{Benfield01},
and sensors for chemical agents~\cite{Novak03}; not to mention more
commonplace applications such as filtering and humidity
sensors~\cite{Varghese02}.  The breadth of nanopore research serves
to underscore the need for a solid physical understanding of the
evolution of liquids in such systems.
Experimental studies have been done on porous network systems, such
as the disordered Vycor~\cite{Gelb98} and M41S silica
materials~{\cite{Neimark01}. Unfortunately, these porous systems
have very complicated network geometries that make it rather
difficult to compare the measurements with simple theoretical
predictions. Additionally, most of these studies are done at low
temperatures that are impractical for all of the applications
described above.  Room temperature measurements on ordered
nanoporous systems with nearly ideal geometry should provide insight
into the physical processes governing liquid behavior in more
general nanoporous systems.

    We describe here \emph{in-situ} small angle x-ray
scattering (SAXS) experiments of the equilibrium wetting and
capillary condensation of a room temperature solvent,
perfluoro-methyl-cyclohexane (PFMC), within nanoporous alumina
(Al$_2$O$_3$, pore diameter $\sim20$~nm).  The anodized alumina
system had an ideal geometry described by a parallel arrangement of
cylindrical nanopores with large aspect ratios $\sim$ 1:5,000, see
Fig.~\ref{fig:sem}.~\cite{Masuda95} Experiments were carried out
within an environmental chamber that allowed precise control of the
amount of solvent condensed within the pores via changes in the
chemical potential, $\Delta \mu$, relative to liquid/vapor
coexistence similar to studies of wetting on flat~\cite{Tidswell91}
and nanostructured surfaces~\cite{Gang05}. Both adsorption and
desorption processes were investigated reproducibly via this
technique.

 Porous alumina membranes were prepared electrochemically
using a two-step anodization technique~\cite{Masuda95, Shin04} and
cleaned as described elsewhere~\cite{Alvine06}. The resultant
alumina membrane consisted of an array of cylindrical parallel pores
running the entire thickness of the membrane and open on both ends.
The macroscopic dimensions of the nanoporous membrane were about
1~cm~$\times$~1~cm~$\times$~90~microns.  Pores formed a 2D local
hexagonal order (see Fig.~\ref{fig:sem}) with nearest neighbor
distances (center to center) of 58$\pm$4~nm, and diameter of
$21\pm5$~nm, determined via electron microscopy.  After cleaning,
samples were dried and loaded into a hermetically sealed
environmental chamber under an atmosphere of ``ultra-pure" grade
N$_2$.
\begin{figure}[tbp]
  \includegraphics[width=1\columnwidth]{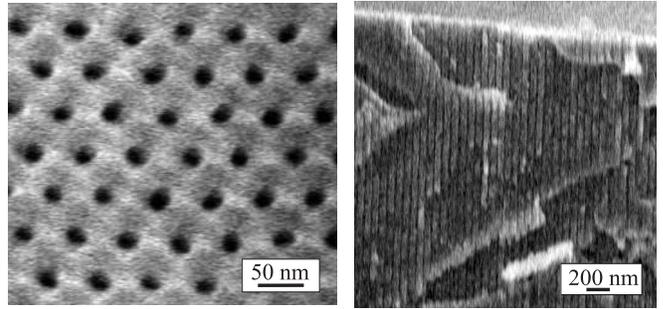}
  \caption{(a) SEM image shows well ordered hexagonal packing of alumina
  nanopores.  (b) SEM of the fractured edge of the membrane showing the
  parallel packing of the nanopore channels (seen as vertical lines) normal to the membrane surface.} \label{fig:sem}
\end{figure}

    The environmental chamber as described elsewhere~\cite{Alvine06,Heilmann01,Tidswell91},
consisted of two concentric cylindrical metal chambers that allowed
extremely stable (drift of $<5$~mK/hr) temperature control of the
sample with a precision of $\pm 1$~mK.  Condensation of liquid
solvent into the pores was precisely controlled via a positive
offset $\Delta T$ between the sample temperature, $T_s$ and the
temperature, $T_r$, of a liquid PFMC (purum grade, 97\% GC)
reservoir, related to the chemical potential by $\Delta\mu\approx
H_{\mathrm{vap}}\Delta T/T_r$.~\cite{Alvine06,Heilmann01,Tidswell91}
Here, $H_{\mathrm{vap}}=33.9$~kJ/mol is the heat of vaporization of
PFMC.~\cite{Good59}   For large $\Delta T$, little or no liquid
condenses in the pores.  As the $\Delta T$ (and thus $\Delta \mu$)
is decreased toward zero, increasing amounts of liquid condense into
the pores until saturation (complete volume filling) is obtained.

    During the experiment, the inner and outer chambers were set at
$T_{in}=31.6$~$^{\circ}$C ($T_r=31.9$~$^{\circ}$C) and
$T_{out}=28.5$~$^{\circ}$C respectively. Although the sample was
nearly thermally isolated from the inner chamber; for high sample
temperatures, such as $T_s=65.4$~$^{\circ}$C, the large heat loads
on the sample increased $T_{in}$ and $T_r$ to 36~$^{\circ}$C and
38~$^{\circ}$C), respectively. At each $\Delta T$ measured however;
the stability of $T_r$ was better than 50~mK.

\begin{figure}[tbp]
  \includegraphics[width=1.0\columnwidth]{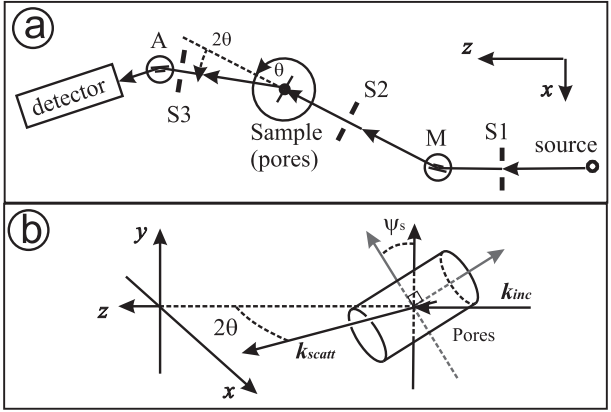}
 \caption{SAXS Geometry:  (a)  Spectrometer setup:  A triple-bounce
 monochromator (M) selects Cu K$_{\alpha 1}$ radiation with high angular
 resolution in the horizontal plane.  An identical triple-bounce analyzer crystal (A) is used after the
 sample.  (b)  Alumina pores are tilted, $\psi_s=8.5^\circ$, in the $y-z$ plane away
 from the parallel condition to take advantage of the enhanced horizontal
 resolution.  Scattered intensity is measured as a function of $2\theta$ in the $x-z$ plane.} \label{fig:geomNA}
\end{figure}

\emph{In-situ} small angle x-ray scattering (SAXS) data were
collected using incident monochromatic Cu-K$_{\alpha 1}$ radiation
with high horizontal angular resolution of $0.005^\circ$
(87~$\mu$rad) obtained using a triple-bounce channel cut
Bonse-Hart~\cite{Deutsch80} Ge(111) monochromator . High horizontal
detector resolution was obtained by mounting an identical
triple-bounce Ge(111) analyzer crystal after the sample, in the
non-dispersive geometry~\cite{Als-Nielsen01} (see
Fig.~\ref{fig:geomNA}). Scattered photons were collected with a
scintillation detector.

 The alumina membrane was mounted in a transmission
geometry with the face of the membrane nearly perpendicular to the
x-ray beam, aligning the long axis of the pores nearly parallel to
the beam. In order to take advantage of the narrow horizontal
angular resolution, the membrane was tilted in the vertical
direction by $\psi_s\approx8.5^\circ$.  This tilt effectively
confined the scattering to the horizontal plane and minimized
wave-guide or multiple scattering effects due to grazing angle
reflections from the pore walls. In this geometry the intensity of
the $\langle 10\rangle$ powder diffraction peak, due to the 2D
hexagonal packing arrangement of the co-aligned pores with a well
defined pore-pore distance, was recorded \emph{in-situ} as $\Delta
T$ was varied between 0.1~K up to 27~K for both adsorption (cooling)
and desorption (heating) paths.

    Immediately prior to collecting the data presented here, the pores were
cleaned in place by repeated ``flushing" (produced by rapid thermal
cycling over the full $\Delta T$ range) with PFMC over ten thermal
cycles. The intensity of the $\langle 10\rangle$ diffraction peak
gradually increased with flushing as compared  at the highest
$\Delta T$; then remained constant after three cycles.

As liquid condenses into the pores, replacing N$_2$, the electron
density contrast of the system decreases, eventually saturating when
the pores are completely filled with liquid.  This reduction in
contrast is measured by a corresponding decrease in the scattering
intensity of the powder diffraction peaks, $I_{peak}(\Delta T)$,
relative to the dry peak intensity, $I_{peak}(dry)$ (see
Fig.~\ref{fig:hysteresis}). Neglecting absorption, the added solvent
volume fraction, $V(\Delta T)$, can be determined within the small
angle approximation~\cite{Alvine06,Gang05} from the lowest order
peak as follows:
\begin{eqnarray}
V(\Delta T)&\propto&\sqrt{I_{peak}(dry)}-\sqrt{I_{peak}(\Delta T)}
\end{eqnarray}
Due to van der Waals (vdW) interactions, the pores are, in practice,
never "dry" or empty making $I_{peak}(dry)$ difficult to measure.
Even at the large $\Delta T$ realized in this experiment we still
expect monolayer adsorption of material onto the walls of the pores,
evidenced by the necessity of ``flushing".  For small $\Delta T$,
complete volume filling is easily observed by the intensity
saturation.  For the purposes of this analysis we calculated a
theoretical value for $I_{peak}(dry)$ from the ratio of the form
factors squared for empty and PFMC filled pores. The form factor, a
function of the radius, $R$, liquid film thickness, $d$, the
electron densities, $\rho_L$ and $\rho_S$, for PFMC and Al$_2$O$_3$
respectively, and scattering vector $q=(2\pi/\lambda)\sin(2\theta)$,
is as follows:
\begin{eqnarray*}
F(\rho_L, \rho_S, d, R)&=&\frac{\rho_L R}{q}\left[R
J_1(qR)-(R-d)J_1(q(R-d))\right]
\end{eqnarray*}
\begin{eqnarray}
 && -\frac{\rho_S R}{q}J_1(qR).
\end{eqnarray}
The ratio, $F(a=0)/F(a=R)\approx0.3$,  combined with the saturated
intensity at low $\Delta T$, provided the necessary offset and
normalization.

Values for the fractional volume condensed into the pores as a
function of $\Delta T$, derived from the $\langle 10\rangle$
diffraction peak intensity for both adsorption and desorption, are
shown in Fig.~\ref{fig:hysteresis}.  X-ray adsorption corrections
were applied to the highest $\Delta T$ points where the heat load
caused increased vapor pressure in the environmental chamber.  The
system exhibited a pronounced capillary transition upon filling,
occurring at approximately $\Delta T^*=1.9\pm0.2$~K (at maximum
slope). Utilizing the simple Kelvin equation for capillary filling:
\begin{eqnarray}
\Delta
T^*=\left(\frac{T_r}{H_{\mathrm{vap}}}\right)\Delta\mu^*=\left(\frac{T_r}{H_{\mathrm{vap}}}\right)\frac{\gamma}{nR}\quad,
\end{eqnarray}
where $\gamma=12.5$~mN/m is the surface tension of the PFMC, and $n=
5.1\times 10^{3}$~mol/m$^3$ is the molar volume of the PFMC, this
indicates a nanopore radius of $R=11.8\pm 1$~nm, in agreement with
electron microscopy measurements.  Here vdW interactions have been
ignored.  The system also exhibited hysteresis common to porous
systems~\cite{Adamson97} with the desorption transition occurring at
$\Delta T^*=2.3\pm0.2$~K. The width of this hysteresis is smaller by
a factor of five than the prediction by Cohan~\cite{Cohan38} for a
desorption transition at $2\Delta T^*$ due to pores emptying by
retreating menisci from the pore ends. Thus, the simple Cohan
argument cannot provide a complete description of the hysteresis
observed here, indicating the necessity of using a more complex
theory.
\begin{figure}[tbp]
  \includegraphics[width=1.0\columnwidth]{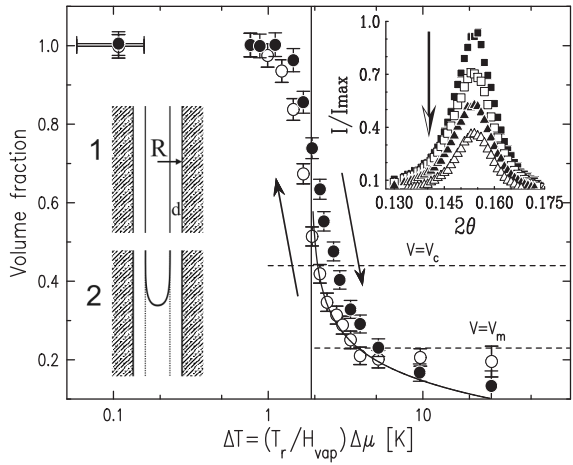}
 \caption{Main:  Volume fraction as a function of $\Delta T$.
 Adsorption, ($\circ$), and desorption, (\textbullet), data are
 shown with arrows indicating the measurement direction.  A vertical
 line represents the Kelvin transition, $\Delta T=1.9\pm0.2$~K.  The solid curve
 is a fit of the pre-transition filling within the Derjaguin approximation.
Horizontal dashed lines indicate the critical and metastable volume
fractions, $V_c$ and $V_m$ respectively. Inset (LEFT):  (1)
 conformal pore filling with film thickness, $d$; (2) partially filled (metastable condition) pore with meniscus
 during desorption. Inset (RIGHT):  The $\langle10\rangle$ diffraction peak
 for various $\Delta T$ during the absorption cycle. The arrow indicates
 the intensity decrease with decreasing $\Delta T$.} \label{fig:hysteresis}
\end{figure}

In their seminal work, Cole and Saam~\cite{Cole74,Saam75} put forth
a theoretical treatment which includes substrate interactions
ignored by the Cohan-Kelvin theory. A brief outline of their
capillary pore filling theory, including the influence of vdW
interactions with the pore walls is given here.  We make an
additional assumption, valid for this system, that the capillary
transition occurs when the adsorbed thickness of liquid on the pore
walls is thin enough, with respect to the pore radius, to employ the
Derjaguin approximation~\cite{Derjaguin76, Israel92}, allowing a
simplified analytical treatment.

The total free energy per unit length of a cylindrically symmetric
film on the walls of the pore is:
\begin{eqnarray}
G(d) = 2\pi\gamma (R-d) + n\Delta\mu\pi (2Rd-d^2) + U(d)\quad,
\end{eqnarray}
where $U(d)$ is the vdW potential between the liquid and the pore
walls.  For thin films relative to the pore radius, $R\gg d$, $U(d)$
is approximately that of a liquid film on a flat substrate
(Derjaguin approximation)~\cite{Israel92}:
\begin{eqnarray}
U(d) = \frac{A_{\mathrm{eff}}}{12\pi d^2}2\pi R
\end{eqnarray}
where $A_{\mathrm{eff}}=6.3\times 10^{-20}$~J is the effective
Hamaker constant for PFMC on Al$_2$O$_3$~\cite{note1}.  In the limit
of vanishing film thickness, this term should be replaced by the
interfacial tension term for the pore wall/liquid interface. The
equilibrium condition is then given by:
\begin{eqnarray}
\frac{\partial G(d)}{\partial N}&=&0\\
\Delta\mu(d) &=&\frac{A_{\mathrm{eff}}}{6\pi n
d^3}+\frac{\gamma}{n(R-d)}\label{vdW}
\end{eqnarray}
Equation~\ref{vdW} simplifies to the Kelvin equation when vdW
interactions are ignored.  Using Eq.(~\ref{vdW}) we fit the
pre-transition filling data for the absorption cycle as a function
of $R$ yielding a value of $R=11.8\pm0.2$ (see
Fig.~\ref{fig:hysteresis} solid curve) in agreement with the Kelvin
value above.  Discrepancies with the fit at large $\Delta T$ are due
to the film thickness being on the order of the monolayer thickness.
The capillary transition to filled pores occurs when the chemical
potential reaches the minimum allowed value, since further reduction
of the chemical potential can only be achieved by complete filling
of the pore, $d=R$, in which case the surface tension term vanishes
due to the vanishing of the interface. This stability or critical
condition for the cylindrically symmetric film is satisfied for
$\partial \Delta\mu(d)/\partial N \geq 0$ (in the limit $d \ll R$):
\begin{eqnarray}
\frac{A_{\mathrm{eff}}}{2\pi d^4}\geq
\frac{\gamma}{(R-d)^2}\approx\frac{\gamma}{R^2}
\end{eqnarray}
This demonstrates that the thin film is stabilized by the vdW
interactions with the pore wall and that, at the transition,
$d\approx3.3$~nm validating the use of the Derjaguin approximation,
$d \ll R$.

    For our system with $R=11.8$~nm (using the fit value), numerical
calculations of Cole and Saam predict an onset of instability for
volume fraction $V_c\approx0.44$ (see Fig. 3 in Ref.~\cite{Saam75})
with $R_0=\sqrt{A_{\mathrm{eff}}/2\gamma}\approx1.6$~nm for the
PFMC/Al$_2$O$_3$ system. The estimate for the onset of the capillary
transition, $V_c$ occurs at slightly higher $\Delta T$ than expected
from the Kelvin and Derjaguin methods (see
Fig.~\ref{fig:hysteresis}), but within the uncertainty given above.

   Cole and Saam additionally make a prediction about the hysteresis
upon desorption.  They argue that the hysteresis occurs due to a
partially filled pore with meniscus (see Fig.~\ref{fig:hysteresis},
upper insert (2)) being metastable with respect to a conformal thin
film geometry (see Fig.~\ref{fig:hysteresis}, upper insert (1)). The
metastable limit is also a function of the pore radius, R, and
$R_0$.  Cole and Saam's numerical analysis predicts a metastable
volume fraction $V_m \approx 0.23$ for our experimental parameters
(see Fig. 3 in Ref.~\cite{Saam75}).  The $V_m$ value approximately
corresponds to the volume fraction where the desorption curve joins
with the absorption curve, in good agreement with our experimental
results, as indicated in Fig.~\ref{fig:hysteresis}.  It is worth
noting that this joining occurs at $\Delta T \approx 2 \Delta T^*$,
where the Cohan argument predicts pores to empty by hemispherical
menisci. Unfortunately there is no prediction of the $\Delta T$
where the desorption transition initiates, and it not possible to
translate $V_c$ and $V_m$ into a hysteresis width.

    The above analysis indicates that for systems with
ideal capillary geometry, such as the porous anodized alumina system
studied here, the theory of Cole and Saam provides a good
description of the capillary filling transition.  For the
cylindrical pores described here, the Kelvin equation is a
reasonable predictor of the filling transition as well, though it
must become increasingly worse for decreasing pore size.  The
Kelvin-Cohan argument fails to qualitatively predict the observed
hysteresis for the system. Cole and Saam's theory, which includes
vdW interactions, demonstrates good agreement with the experimental
results in prediction of the capillary transition and does yield
information about the observed hysteresis, yet there is not a
satisfactory prediction of the observed width.  It is clear from the
experimental data that the desorption transition occurs at lower
$\Delta T$ than either prediction. Some of the shift may be due to
broadening of the transition by pore diameter polydispersity, though
from Fig.~\ref{fig:hysteresis} this effect seems insufficient to
fully account for the shift.  Also, large pore aspect ratios rule
out the influence of end effects and suggests other mechanisms for
desorption are required to explain this feature of the data.

We are grateful to M. W. Cole for helpful discussions and comments.
This work was supported by the National Science Foundation Grants
No. 03-03916 and No. 01-24936


\bibliographystyle{unsrt}

\end{document}